\documentstyle[sprocl]{article}

\def\be{\begin{equation}}
\def\ee{\end{equation}}
\def\bea{\begin{eqnarray}}
\def\eea{\end{eqnarray}}
\def\lqcd{\Lambda_{\rm QCD}}
\def\mev{\,{\rm MeV}}

\def\case#1#2{\textstyle{#1\over#2}}

\begin{document}

\begin{flushright}
JHU--TIPAC--97014\\
hep-ph/9707295\\
July, 1997
\end{flushright}

\vspace{1cm}

\title{THE THEORY OF HEAVY BARYONS~\footnote{To appear in the
Proceedings of  ``$B$ Physics and $CP$ Violation'',
Honolunu, Hawaii, March 24-27, 1997.}}

\author{ADAM F.~FALK}

\address{Department of Physics and Astronomy\\
The Johns Hopkins University\\
3400 North Charles Street\\
Baltimore, Maryland 21218 U.S.A.}

\maketitle\abstracts{I discuss the theory of heavy baryon spectroscopy,
emphasizing predictions which follow from light and heavy quark symmetries.}

The experimental data which are available for the properties of heavy (charmed and
bottom) baryons continue to improve steadily in both quantity and quality.  As a
result, it is important for us to take more seriously the theory of these
states, so that these measurements can be used to further our understanding of
the strong interactions.  In particular, where once we might have been
satisfied with rough predictions derived from quark potential models, now we may
insist on rigorous analyses based on symmetries of QCD.  There has been
significant recent progress in this direction.  In this talk, I review the modern
theory of heavy hadrons, emphasizing the implications of light and heavy quark
symmetries.\cite{Falk}

For a long time, the theory of heavy baryon spectroscopy has been based on quark
potential models, which have at times proven quite successful.  Still, it is
important to keep in mind that quark models are not QCD.  Instead, they are based
on {\it ad hoc,} if often reasonable, assumptions.  Models are often necessary and
sometimes useful, but they are never, despite the success of some of their
predictions, ``correct.''  For example, consider the nonrelativistic constituent
quark model, which is popular for exploring hadron spectroscopy.  The first step
in the construction of this model is to replace the quarks of QCD with new degrees
of freedom, the constituent quarks.  The difference is dramatic; whereas light quarks
have masses of a few MeV and interact via a nonperturbative gauge interaction,
constituent quarks have masses of several hundreds of MeV, interact through a weak
instantaneous potential, and have suppressed magnetic moments.  The second step is
to guess an appropriate potential, the form of which is justified {\it post hoc} by its
success in reproducing the data.  Finally, one calculates the spectrum with these
new degrees of freedom and interactions.  The errors in such an approach are
entirely uncontrolled, as there is no systematic expansion in a small parameter.

As it turns out, the nonrelativistic constituent quark model often works extremely well
for static properties of hadrons, such as spectroscopy.  It is an open
question why this is the case, although there has been interesting
speculation connected with the large $N_c$ limit of QCD.\cite{Manohar}
Another model whose success is hard to understand from fundamental considerations is that of Imbo.\cite{Imbo}  In this model, one assumes two-body interactions between constituent quarks, with an extremely general class of potentials.  The spectrum is then obtained in the limit $D\to\infty$, where $D$ is the number of spatial dimensions!  The trick is to identify relations which are, within the model, as insensitive as possible to the form of the potential.  The result is a collection of predictions which work surprisingly well, typically at the level of a few MeV.  Perhaps it is intriguing to speculate on whether there is a deep reason for this success.

In this talk, we will go beyond models, to identify and exploit {\it
symmetries\/} of QCD and explore their consequences.  Such an approach is less
ambitious than a model-dependent analysis, in the sense that it will lead to fewer concrete
predictions.  However, it is more reliable, and the predictions which are made
come with meaningful estimates of their accuracy.

One interesting class of symmetries arises from the heavy quark limit.\cite{review}  Consider a
hadron composed of a heavy quark $Q$ (bottom or charm) together with some
degrees of freedom consisting of light quarks and gluons.  The light degrees of freedom have energy whose scale is set
by $\lqcd$, the scale of nonperturbative QCD.  In the limit $m_Q\gg\lqcd$, $Q$
acts as a static source of color field, analogous to the proton in a hydrogen
atom.  Since $Q$ does not recoil against the light degrees of freedom, they are
insensitive to $m_Q$, hence to whether $Q=b$ or $Q=c$.  Since the magnetic moment
of $Q$ is suppressed, $\mu_Q\propto 1/m_Q$, they are also insensitive to the
orientation of the spin of $Q$.  The implication is a new heavy quark ``spin-flavor'' $SU(4)$ symmetry, which leads to new conserved quantum numbers of the light
degrees of freedom.  In the heavy
quark limit, the light degrees of freedom may be classified by their excitation
energy $\Delta E$ and their spin-parity $J_\ell^P$.  We will organize heavy
baryon spectroscopy according to this new classification, which is well defined
only in the heavy quark limit of QCD.

There are also symmetries involving the light quarks themselves.  The simplest of these
arises in the limit $m_u=m_d=m_s$, in which case there is a global $SU(3)$ flavor
symmetry.  If $m_u,m_d,m_s\to0$, this symmetry expands to $SU(3)_L\times
SU(3)_R$, a chiral symmetry with implications for the interaction of hadrons with
soft pions.\cite{Georgi}  Finally, if we neglect interactions which flip quark spins, a limit
which is related to $N_c\to\infty$, then we recover the $SU(6)$ which is
represented by the constituent quarks of the nonrelativistic quark model.  In any
of these cases, it is useful to classify the light degrees of freedom by
representations under the symmetries, and to organize the Lagrangian by the symmetry breaking interactions.

\begin{table}
  \caption{The lowest lying charmed baryons.  Isospin is denoted by
$I$, strangeness by $S$.}
  \vspace{0.4cm}
  \centerline{
  \begin{tabular}{|l|l|llllr|l|}
  \hline
  Name&$J^P$&$s_\ell$&$L_\ell$&$J^P_\ell$&$I$&$S$&Decay\\
  \hline
  $\Lambda_c$&$\case12^+$&0&0&$0^+$&0&0&weak\\
  $\Sigma_c$&$\case12^+$&1&0&$1^+$&1&0&
  $\Lambda_c\pi$, $\Lambda_c\gamma$, weak\\
  $\Sigma^*_c$&$\case32^+$&1&0&$1^+$&1&0&$\Lambda_c\pi$\\
  $\Xi_c$&$\case12^+$&0&0&$0^+$&$\case12$&$-1$&weak\\
  $\Xi'_c$&$\case12^+$&1&0&$1^+$&$\case12$&$-1$&$\Xi_c\pi$, $\Xi_c\gamma$\\
  $\Xi^*_c$&$\case32^+$&1&0&$1^+$&$\case12$&$-1$&$\Xi_c\pi$\\
  $\Omega_c$&$\case12^+$&1&0&$1^+$&0&$-2$&weak\\
  $\Omega^*_c$&$\case32^+$&1&0&$1^+$&0&$-2$&$\Omega_c\gamma$\\
  \hline
  \end{tabular}}
  \label{baryontable}
\end{table}

The lowest lying charmed baryons are listed in Table~\ref{baryontable}.  In the second column is the overall $J^P$ of the state, while in the next five columns are given various angular momentum and flavor quantum numbers of the light degrees of freedom.  Note that the distinction between spin $s_\ell$ and orbital $L_\ell$ angular momentum exists only in the quark model, not in QCD, where just the overall $J_\ell^P$ is defined.  The spectrum of bottom baryons is analogous.

States with different $J_\ell^P$ are split by energies of order $\lqcd$, independent of $m_c$.  By contrast, states with the same $J_\ell^P$ but different $J^P$ are split by the chromomagnetic interaction, which scales as $1/m_c$.  Such a pair of states, for example $\Sigma_c$ and $\Sigma_c^*$, are part of a single ``heavy quark doublet.''  It is convenient to define a spin-averaged mass for each doublet, in which the contribution of the chromomagnetic energy is removed.  For the $\Sigma_c$ and $\Sigma_c^*$, this mass is
\be
  \overline\Sigma_c=\textstyle{1\over3}\left(\Sigma_c+2\Sigma_c^*\right)\,,
\ee
where here the states stand for their masses.  We define $\overline\Xi_c$ and $\overline\Omega_c$ analogously.

In the last column of Table~\ref{baryontable}, I list possible decay modes of each baryon.  The dominant decay depends on the available phase space: strong decays will dominate if they are kinematically allowed, followed by radiative decays and finally weak decays.  If $\Sigma_c\to\Lambda_c\pi$ is not allowed, then the decay will also depend on the charge of the $\Sigma_c$:  $\Sigma_c^+$ will decay radiatively, and $\Sigma_c^{0,++}$ weakly.

The symmetries of QCD imply relations between the masses of these baryons.  Neglecting isospin splittings ($m_u\approx m_d$), there are two equal spacing rules,
\bea
  &&\Omega_c-\Xi_c'=\Xi_c'-\Sigma_c\,,\label{su3i}\\
  &&\Sigma_c^*-\Sigma_c=\Xi_c^*-\Xi_c'=\Omega_c^*-\Omega_c\,.\label{su3ii}
\eea
One can use chiral perturbation theory, based on the symmetry $SU(3)_L\times SU(3)_R$, to estimate that the corrections to these relations should be at the level of $10\mev$.\cite{Savage}  There is a third relation which holds in the limit of exact $SU(3)$ symmetry,
\be
  \overline\Sigma_c-\Lambda_c=\overline\Xi_c-\Xi_c\,.\label{su3iii}
\ee
However, this relation receives corrections proportional to $m_s$, which potentially could be quite large.  It is intriguing that the analogous relation in the charmed meson sector holds to within a few MeV.  Finally, one can obtain the same relations (\ref{hqi})--(\ref{hqiii}) using light $SU(6)$.  While the corrections are organized somewhat differently, the conclusions are similar.\cite{JenLeb}

Heavy quark symmetries imply relations between the charmed and bottom hadrons.  The first relates the baryon and meson sectors,
\be\label{hqi}
  \Lambda_b-\Lambda_c=\overline B-\overline D=3340\mev\,,
\ee
where $\overline D={1\over4}(D+3D^*)$.  The second involves the excitation energies of the baryons,
\be\label{hqii}
  \overline\Sigma_b-\Lambda_b=\Sigma_c-\Lambda_c\,,
\ee
and the third the fact that chromomagnetic splittings scale as $1/m_Q$,
\be\label{hqiii}
  (\Sigma_b^*-\Sigma_b)/(\Sigma_c^*-\Sigma_c)=(B^*-B)/(D^*-D)=0.36\,.
\ee
The corrections to the first two relations are of order $\lqcd^2(1/2m_c-1/2m_b)\sim50\mev$; to the third, they are of order $\lqcd^3(1/4m_c^2-1/4m_b^2)$ to each term, or $\sim25\%$ to the ratio.

Note that the $\Xi_c$ and the $\Xi_c'$ have the same overall quantum numbers, but different $J_\ell^P$.  While they are distinct in the heavy quark limit, they can mix when corrections are included.  Such a mixing violates both heavy quark symmetry and $SU(3)$, so it scales as $m_s/m_c$ and is likely to be small.  Estimates based on chiral perturbation theory and quark models give a mixing angle $\theta\le5\%$, so we will neglect this mixing from now on.\cite{mixing}

\begin{table}
  \caption{The observed heavy baryon states, with their conventional
and alternative identities.}
  \vspace{0.4cm}
  \centerline{
  \begin{tabular}{|l|lll|l|l|}
  \hline
  State&Mass (MeV)&Ref.&Decay&Conventional&Alternative \\
  \hline
  $\Lambda_c$&$2285\pm1$&~\cite{PDG}&weak&$\Lambda_c$&$\Lambda_c$\\
  &(2375)&&weak&absent&$\Sigma_c^{0,++}$\\
  &(2375)&&$\Lambda_c+\gamma$&absent&$\Sigma_c^+$\\
  $\Sigma_{c1}$&$2453\pm1$&~\cite{PDG}&$\Lambda_c+\pi$&$\Sigma_c$&
    $\Sigma^*_c$\\
  $\Sigma_{c2}$&$2519\pm2$&~\cite{CLEO96}&$\Lambda_c+\pi$&
    $\Sigma^*_c$&?\\
  $\Xi_c$&$2468\pm2$&~\cite{PDG}&weak&$\Xi_c$&$\Xi_c$\\
  $\Xi_{c1}$&$2563\pm15$\ (?)&~\cite{WA89}
    &$\Xi_c+\gamma$&$\Xi'_c$&$\Xi'_c$\\
  $\Xi_{c2}$&$2644\pm2$&~\cite{CLEO95}&$\Xi_c+\pi$&$\Xi^*_c$&$\Xi^*_c$\\
  $\Omega_c$&$2700\pm3$&~\cite{E687}&weak&$\Omega_c$&$\Omega_c$\\
  $\Omega_c^*$&not yet seen&&&&\\
  \hline
  $\Lambda_b$&$5623\pm6$&~\cite{PDG,CDF96}&weak&$\Lambda_b$&$\Lambda_b$\\
  &(5760)&&weak&absent&$\Sigma_b^\pm$\\
  &(5760)&&$\Lambda_b+\gamma$&absent&$\Sigma_b^0$\\
  $\Sigma_{b1}$&$5796\pm14$&~\cite{DELPHI}&$\Lambda_b+\pi$&
    $\Sigma_b$&$\Sigma^*_b$\\
  $\Sigma_{b2}$&$5852\pm8$&~\cite{DELPHI}&$\Lambda_b+\pi$&
    $\Sigma^*_b$&?\\
  \hline
  \end{tabular}}
   \label{baryondata}
\end{table}

We may now confront the predictions (\ref{su3i})--(\ref{hqiii}) with the available data on the charmed and bottom baryons.  It is important to remember that what is actually determined experimentally is that there is a resonance in a particular decay channel.  The spin and parity of the resonance are not measured; instead, they are usually assigned with the help of a quark model.  (Some additional information may be available from angular distributions of the decay products, but even for weakly decaying states this is never conclusive.)  The observed heavy baryons are listed in Table~\ref{baryondata}, with the mass and the decay channel in which they have been identified.  (Isospin multiplets are averaged over, and, for simplicity, systematic and statistical errors are added in quadrature.)  In the first column I give somewhat arbitrary names to the states, while their conventional (PDG~\cite{PDG}) identities, assigned with quark model input, are given to the right.

Let us compare the states, as conventionally interpreted, with the theoretical predictions.  The light $SU(3)$ relations (\ref{su3i}) and (\ref{su3ii}) may each be rearranged as a prediction for the mass of the $\Xi_c'$.  The first yields $\Xi_c'=2577\mev$ and the second $\Xi_c'=2578\mev$, consistent with each other and with the data.  However the third relation (\ref{su3iii}), when interpreted in the same way, yields $\Xi_c'=2552\mev$, inconsistent with the others.  This disagreement is evidence for large corrections proportional to $m_s$, as compared to those found in the charmed mesons.

The first two heavy quark symmetry prediction also work well, in fact, better than we would have a right to expect.  Experiment yields $\Lambda_b-\Lambda_c=3340\mev$, as compared to $3338\mev$ from theory (\ref{hqi}).  For the second relation (\ref{hqii}), we find $\overline\Sigma_b-\Lambda_b=210\mev$, while $\overline\Sigma_c-\Lambda_c=212\mev$.  However, the spin splitting relation (\ref{hqiii}) fails miserably.  With the measured values of $\Sigma_c$ and $\Sigma_c^*$, we would predict $\Sigma_b^*-\Sigma_b=24\mev$, while the experimental value is 56~\mev.  Although the reported errors are large, this discrepancy is quite significant.

The failed relation is one which really ought to work!  What could be wrong?  Of course, it is possible is that the reported data on $\Sigma_b^{(*)}$ are incorrect; if so, then there is no puzzle to explain.  On the other hand, if the data are confirmed, then one apparently is faced with a dramatic violation of heavy quark symmetry.  Naturally, this is a conclusion which we would be eager to avoid, since such an unexplained failure would call into question many other analyses which are based on this limit of QCD.  The search for a third alternative is strongly motivated.

One possibility is that the conventional assignment of quantum numbers to the observed states is not correct.\cite{Falk}  If some states have been misidentified, then the heavy quark relations have been misapplied, and their failure is no surprise.  Of course, we must be careful not to disturb too badly those predictions which do work well.   The simplest possibility is that the actual $\Sigma_c$ and $\Sigma_b$ states have been missed, because they are too light to decay to $\Lambda_c$ and $\Lambda_b$ by $\pi$ emission.  Rather, they actually decay weakly or radiatively.  In this scenario, the states which conventionally are identified as the $\Sigma_c$ and $\Sigma_b$ are really the $\Sigma_c^*$ and $\Sigma_b^*$, and the conventional $\Sigma_c^*$ and $\Sigma_b^*$ are yet higher resonances.  

If, for example, the undiscovered states had masses $\Sigma_c=2375\mev$ and $\Sigma_b=5760\mev$, then we would have $(\Sigma_b^*-\Sigma_b)/(\Sigma_c^*-\Sigma_c)=0.38$, in perfect agreement with the heavy quark limit.  Furthermore, such an assignment would satisfy the other heavy quark relations within $\sim15\mev$.  The $SU(3)$ relations would now fail by $\sim10\mev$, but this is within the accuracy which we ought to expect for them.  This scenario is illustrated in the last column of Table~\ref{baryondata}; the masses of the proposed states are given in parentheses.

Since the proposed $\Sigma_c$ and $\Sigma_b$ are too light to decay strongly, they would be observed in other channels.  While the $\Sigma_c^+$ would decay radiatively,
\be
  \Sigma_c^+\to\Lambda_c^++\gamma\,,\nonumber
\ee
the others would decay weakly, such as by
\be
  \Sigma_c^{++}\to\left\{\Sigma^++\pi^+\quad{\rm or}\quad\Sigma^++\ell^++\nu
  \quad{\rm or}\quad p+\pi^++K_S\right\},
\ee
and
\be
  \Sigma_c^0\to\left\{\Sigma^-+\pi^+\quad{\rm or}\quad\Sigma^-+\ell^++\nu
  \right\}.
\ee
The decays of the $\Sigma_b$ would be analogous.  The experimental challenge is either to find these states or to rule them out.

In summary, we have seen that heavy quark and light flavor symmetries yield predictions which go beyond the quark model, and which are accompanied by meaningful error estimates.  While the predictions of light $SU(3)$ seem to work well, one of the heavy quark symmetry predictions fails badly with the conventional assignment of quantum numbers to the observed states.  If the data are confirmed, something very interesting is happening, which will have to be understood before heavy quark symmetry can be relied on with confidence in {\it any\/} context.  Until then, we can draw the important moral that it is dangerous to use quark model assumptions to guess the quantum numbers of the observed resonances.  Anything which is not actually measured must be treated as unknown!  As always, the data will tell us what is really going on.

\section*{Acknowledgements}

This work was supported by  the National Science Foundation under
Grant No.~PHY-9404057 and National Young Investigator Award
No.~PHY-9457916; by the Department of Energy under Outstanding
Junior Investigator Award No.~DE-FG02-94ER40869; and by the Alfred
P.~Sloan Foundation.  A.F.~is a Cottrell Scholar of the Research Corporation.

\end{document}